\PassOptionsToPackage{unicode}{hyperref}
\PassOptionsToPackage{hyphens}{url}
\PassOptionsToPackage{dvipsnames,svgnames,x11names}{xcolor}
\documentclass[
  12pt]{article}

\usepackage{amsmath,amssymb}
\usepackage{iftex}
\ifPDFTeX
  \usepackage[T1]{fontenc}
  \usepackage[utf8]{inputenc}
  \usepackage{textcomp} 
\else 
  \usepackage{unicode-math}
  \defaultfontfeatures{Scale=MatchLowercase}
  \defaultfontfeatures[\rmfamily]{Ligatures=TeX,Scale=1}
\fi
\usepackage{lmodern}
\ifPDFTeX\else  
\fi
\IfFileExists{upquote.sty}{\usepackage{upquote}}{}
\IfFileExists{microtype.sty}{
  \usepackage[]{microtype}
  \UseMicrotypeSet[protrusion]{basicmath} 
}{}
\makeatletter
\@ifundefined{KOMAClassName}{
  \IfFileExists{parskip.sty}{%
    \usepackage{parskip}
  }{
    \setlength{\parindent}{0pt}
    \setlength{\parskip}{6pt plus 2pt minus 1pt}}
}{
  \KOMAoptions{parskip=half}}
\makeatother
\usepackage{xcolor}
\makeatletter
\ifx\paragraph\undefined\else
  \let\oldparagraph\paragraph
  \renewcommand{\paragraph}{
    \@ifstar
      \xxxParagraphStar
      \xxxParagraphNoStar
  }
  \newcommand{\xxxParagraphStar}[1]{\oldparagraph*{#1}\mbox{}}
  \newcommand{\xxxParagraphNoStar}[1]{\oldparagraph{#1}\mbox{}}
\fi
\ifx\subparagraph\undefined\else
  \let\oldsubparagraph\subparagraph
  \renewcommand{\subparagraph}{
    \@ifstar
      \xxxSubParagraphStar
      \xxxSubParagraphNoStar
  }
  \newcommand{\xxxSubParagraphStar}[1]{\oldsubparagraph*{#1}\mbox{}}
  \newcommand{\xxxSubParagraphNoStar}[1]{\oldsubparagraph{#1}\mbox{}}
\fi
\makeatother

\usepackage{longtable,booktabs,array}
\usepackage{calc} 
\usepackage{etoolbox}
\makeatletter
\patchcmd\longtable{\par}{\if@noskipsec\mbox{}\fi\par}{}{}
\makeatother
\IfFileExists{footnotehyper.sty}{\usepackage{footnotehyper}}{\usepackage{footnote}}
\makesavenoteenv{longtable}
\usepackage{graphicx}
\makeatletter
\def\maxwidth{\ifdim\Gin@nat@width>\linewidth\linewidth\else\Gin@nat@width\fi}
\def\maxheight{\ifdim\Gin@nat@height>\textheight\textheight\else\Gin@nat@height\fi}
\makeatother
\setkeys{Gin}{width=\maxwidth,height=\maxheight,keepaspectratio}
\makeatletter
\def\fps@figure{htbp}
\makeatother

\makeatletter
\@ifpackageloaded{caption}{}{\usepackage{caption}}
\AtBeginDocument{%
\ifdefined\contentsname
  \renewcommand*\contentsname{Table of contents}
\else
  \newcommand\contentsname{Table of contents}
\fi
\ifdefined\listfigurename
  \renewcommand*\listfigurename{List of Figures}
\else
  \newcommand\listfigurename{List of Figures}
\fi
\ifdefined\listtablename
  \renewcommand*\listtablename{List of Tables}
\else
  \newcommand\listtablename{List of Tables}
\fi
\ifdefined\figurename
  \renewcommand*\figurename{Figure}
\else
  \newcommand\figurename{Figure}
\fi
\ifdefined\tablename
  \renewcommand*\tablename{Table}
\else
  \newcommand\tablename{Table}
\fi
}
\@ifpackageloaded{float}{}{\usepackage{float}}
\floatstyle{ruled}
\@ifundefined{c@chapter}{\newfloat{codelisting}{h}{lop}}{\newfloat{codelisting}{h}{lop}[chapter]}
\floatname{codelisting}{Listing}

\makeatother
\makeatletter
\@ifpackageloaded{caption}{}{\usepackage{caption}}
\@ifpackageloaded{subcaption}{}{\usepackage{subcaption}}
\makeatother

\ifLuaTeX
  \usepackage{selnolig}  
\fi
\usepackage[]{natbib}
\usepackage{bookmark}

\IfFileExists{xurl.sty}{\usepackage{xurl}}{} 
\hypersetup{
  pdftitle={A discrete-time survival model to handle interval-censored covariates, with applications to HIV cohort studies},
  pdfauthor={Avi Kenny; Stephen Olivier; Jianxuan Zang; Jeffrey W. Imai-Eaton; James P. Hughes; Mark J. Siedner},
  pdfkeywords={time-to-event, serostatus, seroconversion, mortality},
  colorlinks=true,
  linkcolor={blue},
  filecolor={Maroon},
  citecolor={Blue},
  urlcolor={Blue},
  pdfcreator={LaTeX via pandoc}}

\newcommand{\anon}{1}

\DeclareMathOperator*{\argmax}{arg\,max}

\newcommand{\midd}{\,|\,}

\begin{document}

\def\spacingset#1{\renewcommand{\baselinestretch}%
{#1}\small\normalsize} \spacingset{1}


\if1\anon
{
  \title{\bf A discrete-time survival model to handle interval-censored covariates, with applications to HIV cohort studies}
  \author{Avi Kenny\thanks{
    The authors gratefully acknowledge the National Institute Of Allergy And Infectious Diseases of the National Institutes of Health [Award Number R37-AI029168], the National Heart, Lung, and Blood Institute of the National Institutes of Health [Award number K24-HL166024], and the Wellcome Trust [Award number Wellcome Strategic Core award: 227167/A/23/Z].}\hspace{.2cm}\\
    Department of Biostatistics and Bioinformatics, Duke University\\
    Global Health Institute, Duke University\\\\
    Stephen Olivier \\
    Africa Health Research Institute\\\\
    Jianxuan Zang \\
    Department of Biostatistics and Bioinformatics, Duke University\\\\
    Jeffrey W. Imai-Eaton \\
    Center for Communicable Disease Dynamics,\\Department of Epidemiology, Harvard T H Chan School of Public Health\\
    MRC Centre for Global Infectious Disease Analysis,\\School of Public Health, Imperial College London\\\\
    James P. Hughes \\
    Department of Biostatistics, University of Washington\\\\
    Mark J. Siedner \\
    Division of Infectious Diseases, Massachusetts General Hospital\\
    Division of Internal Medicine, University of KwaZulu-Natal\\\\
    }
  \maketitle
} \fi

\if0\anon
{
  \bigskip
  \bigskip
  \bigskip
  \begin{center}
    {\LARGE\bf A discrete-time survival model to handle interval-censored covariates, with applications to HIV cohort studies}
\end{center}
  \medskip
} \fi

\bigskip
\begin{abstract}
Methods are lacking to handle the problem of survival analysis in the presence of an interval-censored covariate, specifically the case in which the conditional hazard of the primary event of interest depends on the occurrence of a secondary event, the observation time of which is subject to interval censoring. We propose and study a flexible class of discrete-time parametric survival models that handle the censoring problem through simultaneous modeling of the interval-censored secondary event, the outcome, and the censoring mechanism. We apply this model to the research question that motivated the methodology, estimating the effect of HIV status on all-cause mortality in a prospective cohort study in South Africa. Our model has applicability for many open questions, including estimating the impact of policy decisions on population level HIV-related outcomes and determining causes of morbidity and mortality for which the HIV positive population may be at increased risk. Examples include determining how the large-scale transition from efavirenz-based to dolutegravir-based first-line ART impacted mortality for people living with HIV and determining whether HIV status is associated with increased risk of stroke, diabetes, hypertension, and other non-communicable diseases.
\end{abstract}

\noindent%
\vfill

\newpage
\spacingset{1.8} 

\section{Introduction}\label{sec-intro}

Survival data, also known as time-to-event data, are ubiquitous in public health research and require specialized methods to handle event times that may not be observed directly, but are instead subject to censoring. Typically, the censored event time is the outcome of interest in an analysis, such as mortality or the onset of a disease. However, in certain situations we may also have a censored covariate (the \textit{secondary event}), related to the outcome (the \textit{primary event}) in the sense that at any point in time, the conditional hazard of the primary event depends on whether the secondary event has occurred.

In particular, we focus on the case in which the outcome of interest is right-censored and a covariate is interval-censored. With \textit{interval censoring}, the value of a covariate $X$ is not known exactly, but is known to lie within an interval $[l,u]$. This definition is broad, and includes right-censoring (in which $u=\infty$ and $X \ge l$), left-censoring (in which $l=-\infty$ and $X \le u$), and what we refer to as ``finite interval censoring'' (in which $l$ and $u$ are both finite and in the interior of the support of $X$). Furthermore, missing data can be thought of as a case of interval-censoring in which $l=-\infty$ and $u=\infty$. Thus, the framework of interval censoring is broad and applicable to a wide variety of settings.

A vast body of research has been developed for handling right-censored outcomes, including classical methods such as the Kaplan-Meier estimator \citep{kaplan1958nonparametric} and the Cox proportional hazards model \citep{cox1972regression}. Additionally, a number of methods are available for when the outcome variable is interval-censored (see, for example, \citealp{lindsey1998methods,kor2013method,pan2002estimation}), including the case when the censoring mechanism is informative \citep{finkelstein2002analysis}; see \cite{gomez2009tutorial} for a review. However, relatively little work has been done to handle situations in which covariates are right-censored or interval-censored; the setting of a survival model in the presence of an interval-censored covariate is the focus of this paper.

This work is motivated by the problem of estimating the effects of HIV serostatus (the interval-censored covariate) on downstream outcomes, such as all-cause mortality or occurrence of a stroke. With the transition of the global HIV epidemic from a routinely fatal disease to a chronic infection with increasing life expectancy, there is growing interest in measuring the effects of chronic HIV infection and its treatment on risk of non-communicable diseases and excess mortality arising from myriad AIDS-related and non-AIDS-related causes. Numerous longitudinal cohort studies, initially designed to monitor the epidemiology of HIV incidence or shorter-term AIDS related complications, have expanded focus to include study of non-AIDS complications and longer-term mortality \citep{d2021characteristics,gange2007cohort,reniers2016data}. This typically requires comprehensively classifying the HIV serostatus of all individuals in these cohorts, including those whose most recent test was negative and those without testing data. Ideally, population-based studies would include routine and updated HIV testing data on all individuals (e.g. annually) to enable such classification. However, in reality, most population-based studies have episodic HIV testing data, which includes participants who lack HIV testing data, irregular HIV testing schedules, and numbers of individuals who have a final HIV test years before chronic disease and mortality outcomes. Individuals in these types of studies fall into one of four categories: (1) those who have a negative HIV test followed by a positive test, (2) those whose most recent test was negative (but may have acquired infection during the period since the most recent test), (3) those whose first test was positive, and (4) those who have never received a test. With respect to the date of seroconversion, these four cases correspond to finite interval censoring, right-censoring, left-censoring, and missing data, respectively. The first of these four types has been studied to an extent in the HIV modeling literature, with simple methods being proposed, such as imputing a random point between the endpoints \citep{vandormael2018incidence,ghys2007survival}; the other cases have received less methodological attention. Sometimes, ad-hoc methods are used in practice, such as censoring HIV-negative individuals at a fixed point (e.g., one year) following the most recent negative test \citep{sonnenberg2005soon, glynn2010high}, censoring all individuals prior to their first HIV test \citep{sonnenberg2004hiv}, imputation of seroconversion dates for individuals whose first test was positive \citep{isingo2007survival}, or censoring HIV-negative individuals after a certain (age and sex specific) period of time following their last negative test \citep{reniers2014mortality}. However, these ad-hoc methods can introduce bias and often discard substantial proportions of observed person-time among HIV-negative individuals. A principled method that can handle all four types of interval censoring simultaneously would enable new lines of research, including estimating the impact of policy decisions on HIV-related outcomes at the population level, and determining causes of morbidity and mortality for which the HIV-positive population may be at increased risk. An example of the former is determining how the large-scale transition from efavirenz-based to dolutegravir-based first-line ART impacted morbidity and mortality for people living with HIV. An example of the latter is determining whether HIV status is associated with increased risk of stroke, diabetes, hypertension, and other non-communicable diseases.

In terms of related works, we identified seven methodological papers that deal with interval-censored covariates in the context of a survival model. \cite{lee2003proportional} and \cite{atem2019cox} consider an adaptation to the Cox proportional hazards model to a setting in which both the outcome and a covariate are right-censored, and \cite{sattar2012parametric} consider a similar setting in which the covariate is instead left-censored and propose a full parametric model. However, none of these methods are appropriate for an interval-censored covariate. The remaining four papers consider an interval-censored covariate. \cite{goggins1999applying} considers a survival data setting similar to our own in which one covariate (the status of a binary event process, measured through periodic tests) is interval censored. They take an approach that involves an EM algorithm with a Gibbs sampling E-step; although this setting is similar to our own, the proposed method is computationally impractical for large datasets, and the imputation of the censored covariate is not allowed to depend on individual-level covariates, a major limitation. \cite{langohr2004parametric} consider HIV status as an interval-censored exposure in a survival model, but the simple log-linear parametric model they propose is too inflexible for our setting (e.g., it does not allow for time-varying covariates) and the distribution of the interval-censored variable is similarly not allowed to depend on individual-level covariates. \cite{tian2006analysis} consider a survival setting in which a binary covariate process is observed at a single time point; their method is useful but not applicable in our setting where individuals may have zero, one, or multiple measurements of the covariate process. The setting of \cite{ahn2018cox} is similar to ours, and they consider three possible modifications to a Cox model. However, their estimators are mainly useful for settings in which the covariate is known to lie between two time points, as they essentially discard all information from person-time intervals following the last observation of the covariate process. Finally, although not framed as a methodological paper, the supplement of \cite{risher2021age} gives a detailed approach for handling a setting involving an interval-censored covariate through construction of a custom Bayesian likelihood model.

The contribution of this work is to propose and study a flexible class of parametric discrete-time survival models that are computationally tractable when applied to large datasets that involve an interval-censored covariate and time-varying covariates. This class of models is broad, and allows for both the conditional hazard of the secondary event and the censoring mechanism to depend on covariates. To our knowledge, this is the only work to propose a solution to inference in a general discrete-time survival model with an interval-censored covariate.

The organization of the remainder of this paper is as follows. In section \ref{sec_methods}, we introduce the data structure and describe our statistical model. In section \ref{sec_simulation}, we conduct a simulation study to evaluate the operating characteristics of our model and confirm code functionality. In section \ref{sec_analysis}, we demonstrate the use of our model through application to a dataset from a large HIV cohort in South Africa. In section \ref{sec_discussion}, we summarize findings, limitations, and future research directions.

\section{Methods}\label{sec_methods}

\subsection{Data structure, ideal model, and parameters of interest}

We begin by describing an ideal data structure that involves no censoring or missingness, and then use this structure to describe the data we observe in reality. Suppose that we have a (possibly open) longitudinal cohort of individuals, indexed by $i \in \{1,...,n\}$, with observations occurring within an \textit{observation window} defined by time (calendar time, in our example), which is discretized into intervals indexed by $j \in \{1,...,J\}$. For each individual, we have observations corresponding to some subset $(s_i,s_i+1,...,t_i) \subset (1,2,...,J)$, which we refer to as the observation interval for that individual. The start time $s_i$ corresponds to either the start date of the observation window ($s_i=1$) or the time at which the subject enters the risk set. The end time $t_i$ represents either the time at which the outcome of interest occurred, the end of the observation window, or the time at which the subject exits the risk set. For an individual $i$ at time $j \in (s_i,s_i+1,...,t_i)$, we observe an outcome indicator $Y_{i,j}$ (e.g., death), a fully-observed covariate vector $\textbf{Z}_{i,j}$ (e.g., age and sex), and an indicator $X_{i,j}$ representing whether a secondary event (e.g., HIV seroconversion) has occurred (where in reality $X_{i,j}$ may be missing). For ease of exposition, we assume that $X_{i,j}$ is univariate (i.e., there is only one secondary event); in section \ref{sec_discussion}, we discuss possible extensions to handling multivariate $X_{i,j}$. We also assume that for each individual, the primary outcome can only occur once, and that observation stops for that individual once it has occurred. Also let $\textbf{X}_i \equiv (X_{i,s_i},X_{i,s_i+1},...,X_{i,t_i})$ and $\textbf{X} \equiv (\textbf{X}_1,\textbf{X}_2,...,\textbf{X}_n)$, and define $\textbf{Y}_i$, $\textbf{Z}_i$, $\textbf{Y}$, and $\textbf{Z}$ analogously.

For each individual, we assume that the data follow a longitudinal process in which two Markov-type assumptions hold. First, we assume that the probability (discrete hazard) of the secondary event $X_{i,j}$ occurring at time $j$ depends only on the fully observed covariates $\textbf{Z}_{i,j}$ at time $j$, and does not depend at all on the outcome history. Formally, this can be written as

\begin{equation}\label{eq_markov_1}
    P(X_{i,j}=\textbf{x} \midd
    \bar{X}_{i,j-1}, \bar{Y}_{i,j-1}, \bar{\textbf{Z}}_{i,j})
    = P(X_{i,j}=\textbf{x} \midd \textbf{Z}_{i,j}, X_{i,j-1}=0) \,,
\end{equation}

\noindent where $\bar{\textbf{Z}}_{i,t} \equiv (\textbf{Z}_{i,s_i},\textbf{Z}_{i,s_i+1},...,\textbf{Z}_{i,t})$ represents the covariate history of $\textbf{Z}$ up to time $t$, and analogous definitions hold for $\bar{\textbf{X}}_{i,t}$ and $\bar{\textbf{Y}}_{i,t}$. Second, we assume that the probability (discrete hazard) of the outcome occurring at time $j$ depends only on whether or not it occurred at time $j-1$ and the covariates (including the secondary event process) at time $j$. This can be written as

\begin{equation}\label{eq_markov_2}
    P(Y_{i,j}=\textbf{y} \midd
    \bar{X}_{i,j}, \bar{Y}_{i,j-1}, \bar{\textbf{Z}}_{i,j})
    = P(Y_{i,j}=\textbf{y} \midd X_{i,j}, Y_{i,j-1}, \textbf{Z}_{i,j}) \,.
\end{equation}

\noindent Note that both probabilities are allowed to depend on study time $j$. Next, we define $p_x^*$ as $\Omega$

\begin{equation*}
    p_x^*(\tilde{x},z,j,s,\Omega) \equiv P(X_{i,j}=1 \midd X_{i,j-1}=\tilde{x}, \textbf{Z}_{i,j}=\textbf{z}, j, I(j=s_i)=s, \Omega) \,,
\end{equation*}

\noindent the conditional probability that the secondary event has occurred by time $j$ given $(X_{i,j-1}, \textbf{Z}_{i,j}, I(j=s_i))$. Note that $p_x^*(\tilde{x}=1,\textbf{z},j,s,\Omega)=1$ for all $(\textbf{z},j,s,\Omega)$ and that $p_x^*(\tilde{x}=0,\textbf{z},j,s,\Omega)$ is a discrete hazard function. Also note that $X_{i,s_i-1}$ is undefined, and a model for $p_x^*$ must account for this ``initial status''; this is the reason for conditioning on $I(j=s_i)$, an indicator that the current observation is the first for individual $i$. We also denote by $p_y^*(x,\textbf{z},j,\Omega)$ the discrete hazard (probability) of the outcome occurring in time interval $j$ given $(X_{i,j}=x, \textbf{Z}_{i,j}=\textbf{z}, \Omega)$; that is,

\begin{equation*}
    p_y^*(x,\textbf{z},j,\Omega) \equiv P(Y_{i,j}=1 \midd Y_{i,j-1}=0, X_{i,j}=x, \textbf{Z}_{i,j}=\textbf{z}, j, \Omega) \,,
\end{equation*}

\noindent The corresponding conditional probability mass functions (PMFs) are given by

\begin{align*}
    f_x^*(x \midd \tilde{x},\textbf{z},j,s,\Omega) &\equiv \{p_x^*(\tilde{x},\textbf{z},j,s,\Omega)\}^x \{1-p_x^*(\tilde{x},\textbf{z},j,s,\Omega)\}^{1-x} \,, \\
    f_y^*(y \midd x,\textbf{z},j,\Omega) &\equiv \{p_y^*(x,\textbf{z},j,\Omega)\}^y \{1-p_y^*(x,\textbf{z},j,\Omega)\}^{1-y} \,.
\end{align*}

We assume that interest lies in contrasts or parameters related to the discrete hazard function $p_y^*$. For example, if we assume that $p_y^*(x,\textbf{z},j,\Omega) = \mathcal{C}^{-1}(\alpha x + \beta'\textbf{z} + b(j))$, where $\mathcal{C} : x \mapsto \log\{-\log(1-x)\}$ is the complementary log-log link function, $\mathcal{C}^{-1}$ is its inverse, and $b(j)$ is an arbitrary function of calendar time, then the (exponentiated) parameters $\alpha$ and $\beta$ can be interpreted as hazard ratios \citep{prentice1978regression}. Conditions \eqref{eq_markov_1} and \eqref{eq_markov_2} allow us to write the joint PMF of $(\textbf{X}_i,\textbf{Y}_i)$ evaluated at the vector $(\textbf{x},\textbf{y}) \equiv (x_{s_i},...,x_{t_i},y_{s_i},...,y_{t_i})$ given $\textbf{Z}_i = \textbf{z} \equiv (\textbf{z}_{s_i},...,\textbf{z}_{t_i})$ and $\Omega$ as

\begin{equation*}
    f_{x,y}(x,y \midd \textbf{z}, \Omega) \equiv
    \prod_{j={s_i}}^{t_i}
    f_x^*(x_j \midd x_{j-1}, \textbf{z}_j, j, I(j=s_i), \Omega)
    f_y^*(y_j \midd x_j,\textbf{z}_j,j,\Omega) \,.
\end{equation*}

\subsection{Handling the interval-censored covariate process}

\noindent If it were the case that $\textbf{X}$ was fully-observed, inference for $\Omega$ could be based on the log likelihood

\begin{align*}
    \tilde{\ell}_{x,y}(\Omega | \textbf{X},\textbf{Y},\textbf{Z})
    &= \sum_{i=1}^n \log \left\{
        f_{x,y}(\textbf{X}_i,\textbf{Y}_i \midd \textbf{Z}_i, \Omega)
    \right\} \\
    &= \sum_{i=1}^n \sum_{j={s_i}}^{t_i} \log \left\{
        f_x^*(X_{i,j} \midd X_{i,j-1}, \textbf{Z}_{i,j}, j, I(j=s_i), \Omega)
        f_y^*(Y_{i,j} \midd X_{i,j},\textbf{Z}_{i,j},j,\Omega)
    \right\} \,,
\end{align*}

However, in our case, $\textbf{X}$ is partially or completely interval-censored. Instead of observing $X_{i,j}$, we observe $U_{i,j} \equiv \Delta_{i,j} X_{i,j}$, where $\Delta_{i,j}$ is a binary variable that equals one if $X_{i,j}$ is known and zero otherwise. We proceed by (1) positing a mechanism or model for the missingness indicator $\Delta$, (2) deriving the form of the joint PMF of $(\textbf{X},\textbf{Y},\Delta,\textbf{U} \midd \textbf{Z}, \Omega)$, and (3) integrating out $\textbf{X}$ to derive the joint PMF of $(\textbf{Y},\Delta,\textbf{U} \midd \textbf{Z}, \Omega)$, on which inference can be based.

For the first step, the joint PMF of $(\textbf{X}_i,\textbf{Y}_i,\Delta_i)$ given $(\textbf{Z}_i,\Omega)$ for an individual can be written as

\begin{equation*}
    f_{x,y,\delta}(\textbf{x},\textbf{y},\delta \midd \textbf{z}, \Omega) =
    f_{x,y}(\textbf{x},\textbf{y} \midd \textbf{z}, \Omega)
    f_\delta(\delta \midd \textbf{x}, \textbf{y}, \textbf{z}, \Omega) \,,
\end{equation*}

\noindent where $f_\delta$ represents the PMF of the vector $\Delta_i$ conditional on $(\textbf{X}_i,\textbf{Y}_i,\textbf{Z}_i,\Omega)$, the form of which is chosen by the user based on context. Assumptions about the missingness mechanism are implicitly encoded by the form of $f_\delta$. If we assume that $\textbf{X}_i$ is missing completely at random (MCAR), $f_\delta$ would be constant. If $\textbf{X}_i$ is missing at random (MAR), $f_\delta$ would not depend on $\textbf{X}_i$ but may depend on $(\textbf{Y}_i,\textbf{Z}_i)$. If we assume that $\textbf{X}_i$ is missing-not-at-random (MNAR) then $f_\delta$ depends on $\textbf{X}_i$; see section \ref{ssec_model_spec} for an example of one such mechanism.

For the second step, we note that since $\textbf{U}$ equals the product of $\textbf{X}$ and $\Delta$, we can write the conditional PMF of $(\textbf{U}_i,\textbf{X}_i,\textbf{Y}_i,\Delta_i)$ given $(\textbf{Z}_i,\Omega)$ as

\begin{equation*}
    f_{u,x,y,\delta}(\textbf{u},\textbf{x},\textbf{y},\delta \midd \textbf{z}, \Omega) =
    f_{x,y,\delta}(\textbf{x},\textbf{y},\delta \midd \textbf{z}, \Omega) I(\textbf{u}=\textbf{x}\delta) \,.
\end{equation*}

\noindent For the third step, we can obtain the conditional PMF of $(\textbf{U}_i,\textbf{Y}_i,\Delta_i)$ given $(\textbf{Z}_i,\Omega)$ as

\begin{equation}\label{eq_f_uyd}
    f_{u,y,\delta}(\textbf{u},\textbf{y},\delta \midd \textbf{z}, \Omega) =
    \sum_{x\in\mathcal{X}} f_{u,x,y,\delta}(\textbf{u},\textbf{x},\textbf{y},\delta \midd \textbf{z}, \Omega) \,,
\end{equation}

\noindent where $\mathcal{X}$ is the support of $\textbf{X}_i$. Since $\textbf{X}_i$ will always equal a binary vector of the form $(0,...,0,1,...,1)$ (possibly with only zeros or only ones), the set $\mathcal{X}$ can be written as $\mathcal{X} \equiv \{ \textbf{R}_0(k), (\textbf{R}_0(k-1),\textbf{R}_1(1)), ..., (\textbf{R}_0(1),\textbf{R}_1(k-1)), \textbf{R}_1(k) \}$, where $k$ is the length of $\textbf{X}_i$, $\textbf{R}_0(n)$ represents a vector of $n$ zeros, and $\textbf{R}_1(n)$ represents a vector of $n$ ones. The resulting conditional log-likelihood function across all individuals is given by

\begin{equation}\label{eq_likelihood_miss}
    \ell_{u,y,\delta}(\Omega | \textbf{U},\textbf{Y},\Delta,\textbf{Z}) = \sum_{i=1}^n \log \left\{
        f_{u,y,\delta}(\textbf{U}_i,\textbf{Y}_i,\Delta_i \midd \textbf{Z}_i, \Omega)
    \right\} \,.
\end{equation}

\noindent While $\textbf{X}$ is not observed for everyone, the quantity $\textbf{U}$ is always observed, and so inference for $\Omega$ can be based on \eqref{eq_likelihood_miss}. In most applications, the log-likelihood will have to be maximized numerically to compute estimators $\widehat{\Omega}$ and $\widehat{\text{Var}}(\widehat{\Omega})$ of $\Omega$ and $\text{Var}(\hat{\Omega})$, respectively.

\subsection{Specification of model components}\label{ssec_model_spec}

The form of the log-likelihood given in \eqref{eq_likelihood_miss} requires specification of parametric forms for the functions $p_x^*$, $p_y^*$, and $f_\delta$. For the discrete hazard function $p_y^*$, a traditional approach is to use a linear predictor with a complementary log-log link function. This is a discrete analog of a Cox model for continuous-time data, and so the resulting parameters can be interpreted as hazard ratios \citep{prentice1978regression}. For example, we may have that

\begin{equation}\label{eq_p_outcome}
    p_y^*(x,z,j,\Omega) = \mathcal{C}^{-1} \left\{
        \alpha_y + \beta x + \gamma_y'\textbf{z} + \tau_y'\textbf{b}_y(j)
    \right\} \,,
\end{equation}

\noindent where again $\mathcal{C}^{-1}$ is the inverse of the complementary log-log link function and where $\textbf{b}_y$ is a vector-valued basis function (e.g., a natural cubic spline basis) modeling the calendar time trend. For the conditional distribution of the interval-censored variable $\textbf{X}$, a number of models are possible depending and one should be chosen based on context. In the motivating example for this work, the interval-censored covariate is HIV serostatus, a binary vector of the form $(0,...,0,1,...,1)$, with the change from $0$ to $1$ occurring at the point of seroconversion. One choice is to model this as

\begin{align}\label{eq_p_seroconv}
\begin{split}
    p_x^*(\tilde{x},\textbf{z},j,s,\Omega)
    &\equiv \left[
        \tilde{\mathcal{C}}^{-1} \left\{
            \alpha_s + \gamma_s'\textbf{z} + \tau_s'\textbf{b}_s(j)
        \right\}
    \right]^s \\
    &\qquad\times \left[
        \tilde{x} + (1-\tilde{x}) \, \tilde{\mathcal{C}}^{-1} \left\{
            \alpha_x + \gamma_x'\textbf{z} + \tau_x'\textbf{b}_x(j)
        \right\} \,,
    \right]^{1-s} \,,
\end{split}
\end{align}

\noindent where the notation $\tilde{\mathcal{C}}$ is used to emphasize that a different link function can be used if desired. Note that the form given in \eqref{eq_p_seroconv} includes two components: a parametric form for the ``initial status model'' (i.e., the conditional probability that the event has already happened for individual $i$ by the time of the first measurement $j=s_i$) and a separate parametric form for the ``secondary event discrete hazard model'' (i.e., for subsequent measurements at times $j>s_i$, the conditional probability of the event happening in one time interval given that it has not happened up to that point).

The function $f_\delta$ can be used to model the conditional interval censoring mechanism. Different forms are possible, but it will often be the case that measurements of the secondary event will be taken at specific points in time that inform knowledge of the event process at other points in time. In these cases, it is often more convenient to think of $\Delta_{i,j}$ as bivariate, given by $\Delta_{i,j} \equiv (\Delta_{i,j}^\circ,\Delta_{i,j}^*)$, where $\Delta_{i,j}^\circ$ represents whether $X_{i,j}$ is known (as before) and $\Delta_{i,j}^*$ is an indicator that equals one if a test or measurement for individual $i$ was taken at time $j$. Then, we can model the random variable $\Delta^*$ and use this variable to deterministically infer $\Delta^\circ$. For example, with HIV testing, if an individual is only tested once at time $t^*$, then it will be the case that $\Delta_{i,j}^*=1$ for $j=t^*$ and $\Delta_{i,j}^*=0$ for $j\ne t^*$. However, if the test is negative, then we know the event has not occurred at any time prior to $t^*$, and so we have that $\Delta_{i,j}^\circ=1$ for $j \le t^*$ and $\Delta_{i,j}^\circ=0$ for $j > t^*$. This can be represented by defining a conditional PMF function for $\Delta_{i,j}^*$; for example, if we assume this variable depends on the covariate vector $\textbf{Z}_{i,j}$, we may specify the form

\begin{equation}\label{eq_f_deltastar}
    f_{\delta^*}(\delta^* \midd \textbf{z}, \Omega)
    \equiv \prod_{j=s_i}^{t_i}
    \left\{ \tilde{\mathcal{C}}^{-1}(\alpha_\delta+\gamma_\delta'\textbf{z}_j+\tau_\delta'\textbf{b}_{\delta^*}(j)) \right\}^{\delta_j^*}
    \left\{ 1 - \tilde{\mathcal{C}}^{-1}(\alpha_\delta+\gamma_\delta'\textbf{z}_j+\tau_\delta'\textbf{b}_{\delta^*}(j)) \right\}^{1-\delta_j^*} \,.
\end{equation}

\noindent We can then write

\begin{equation*}
    f_{\delta}(\delta \midd \textbf{x}, \textbf{y}, \textbf{z}, \Omega) \equiv
    I\{ \delta^\circ = g_\delta(\textbf{x},\delta^*) \} f_{\delta^*}(\delta^* \midd \textbf{z}, \Omega) \,,
\end{equation*}

\noindent where the function $g_\delta$ computes the vector $\Delta_i^\circ$ as a function of the vector $\Delta_i^*$ of testing times and the true event indicators $\textbf{X}_i$. In the context of our motivating HIV example, this function encodes the idea that (1) HIV status is known to be negative for all times prior to a negative test, and (2) HIV status is known to be positive for all times following a positive test. To give a form for this function, we first define a categorical variable $M_i$ that subdivides the population into four mutually exclusive and exhaustive groups. Group 1 consists of individuals who have no testing data whatsoever. Group 2 consists of individuals who have only ever received negative tests. Group 3 consists of individuals who received one or more negative tests followed by a positive test. Group 4 consists of individuals whose first and only test was positive. Mathematically, this is summarized through the following function, where $x$ and $\delta^*$ are both vectors of length $J$.

\begin{align*}
    m(x,\delta^*) \equiv
    \begin{cases}
        1, & \text{if}\
        \sum_{j=1}^J \delta_j^* = 0 \,, \\
        2, & \text{if}\
        \sum_{j=1}^J \delta_j^* > 0,
        \sum_{j=1}^J \delta_j^* x_j = 0 \,, \\
        3, & \text{if}\
        \sum_{j=1}^J \delta_j^* > 0,
        \sum_{j=1}^J \delta_j^* x_j = 1,
        \sum_{j=1}^J \delta_j^* (1-x_j) > 0 \,, \\
        4, & \text{if}\
        \sum_{j=1}^J \delta_j^* = 1,
        \sum_{j=1}^J \delta_j^* x_j = 1 \,.
    \end{cases}
\end{align*}

\noindent Using this definition, for each individual $i$, let $M_i \equiv m(\textbf{X}_i,\Delta_i^*)$. Also let $T_i^- \in \{1,2,...,J\}$ denote the index of the most recent negative test (defined for groups 2 and 3) and let $T_i^+ \in \{1,2,...,J\}$ denote the index of the (only) positive test (defined for groups 3 and 4); these can be formally defined as $T_i^- \equiv t^-(\textbf{X}_i,\Delta_i^*)$ and $T_i^+ \equiv t^+(\textbf{X}_i,\Delta_i^*)$, where

\begin{align*}
    t^-(x,\delta^*) &\equiv \argmax_{j\in\{1,2,...,J\}} \left\{ j \delta_j^* (1-x_j) \right\} \,, \\
    t^+(x,\delta^*) &\equiv \argmax_{j\in\{1,2,...,J\}} \left\{ \delta_j^* x_j \right\} \,.
\end{align*}

\noindent Recalling that $\textbf{R}_0(n)$ represents a vector of $n$ zeros and $\textbf{R}_1(n)$ represents a vector of $n$ ones, we can calculate $\Delta_i^\circ$ as

\begin{equation}\label{eq_g_delta}
    \Delta_i^\circ = g_\delta(\textbf{X}_i,\Delta_i^*) \equiv
    g_\delta^*(m(\textbf{X}_i,\Delta_i^*), t^-(\textbf{X}_i,\Delta_i^*), t^+(\textbf{X}_i,\Delta_i^*)) \,.
\end{equation}

\noindent where

\begin{align*}
    g_\delta^*(m,t^-,t^+) \equiv
    \begin{cases}
        \textbf{R}_0(J), & \text{if}\ m=1 \,, \\
        \{ \textbf{R}_1(t^-), \textbf{R}_0(J -t^-) \}, & \text{if}\ m=2 \,, \\
        \{ \textbf{R}_1(t^-), \textbf{R}_0(t^+ -t^- -1), \textbf{R}_1(J -t^+ +1) \}, & \text{if}\ m=3 \,, \\
        \{ \textbf{R}_0(t^+ -1), \textbf{R}_1(J -t^+ +1) \}, & \text{if}\ m=4 \,.
    \end{cases}
\end{align*}

Note that the dependence of $\Delta_i^\circ$ on $\textbf{X}_i$ implies that $\textbf{X}_i$ is missing-not-at-random (MNAR). That is, whether or not $X_{i,j}$ is known (i.e., whether or not $\Delta_{i,j}^\circ=1$) for a particular value of $j$ depends on the vector $X_{i,j}$ itself. This is precisely why it is far more convenient to model the conditional distribution of $\Delta_i^*$ rather than that of $\Delta_i^\circ$.

As an alternative approach, it is sometimes convenient to think of the missingness variable as included in the fully-observed covariate vector $\textbf{Z}$. Similarly, the missingness variable can sometimes be computed as a function of one or more variables in $\textbf{Z}$. In these cases, we can write

\begin{equation*}
    f_\delta(\delta \midd \textbf{x}, \textbf{y}, \textbf{z}, \Omega) = I(\delta = \tilde{g}_\delta(z)) \,,
\end{equation*}

\noindent for some function $\tilde{g}_\delta$. Notably, this implies that the models for $f_x^*$ and $f_y^*$ can depend on the missingness variable $\Delta$ (through $\textbf{Z}$) and must account for this dependence accordingly, if it is assumed to exist. In either case, equation \eqref{eq_f_uyd} can be rewritten as

\begin{equation*}
    f_{u,y,\delta}(\textbf{u},\textbf{y},\delta \midd \textbf{z}, \Omega) =
    \sum_{x\in\mathcal{X}} f_{x,y}(\textbf{x},\textbf{y} \midd \textbf{z}, \Omega) f_{\delta^*}(\delta^* \midd \textbf{z}, \Omega)
    I\{ \delta^\circ = g_\delta(\textbf{x},\delta^*) \} I(\textbf{u}=\textbf{x}\delta^\circ) \,.
\end{equation*}

\noindent Plugging this into \eqref{eq_likelihood_miss} provides a basis for inference.

\section{Simulation study}\label{sec_simulation}

We conducted a simulation study to evaluate the operating characteristics of our model and confirm code functionality. Data were generated according to a discrete-time survival process mimicking a simple HIV open cohort dataset, which involved looping over both individuals ($i$) and over time ($j$), where one time interval represents one year. For each individual, a start time $s_i$ between 1 and 20 was sampled uniformly and two baseline covariates were generated, a binary covariate $Z_{i,s_i,1}$ (representing sex) and a uniformly distributed continuous covariate $Z_{i,s_i,2}$, scaled to lie in the interval $[13,60]$ (representing baseline age). At each time point $j\ge s_i$, the following procedure was used to sample $(\textbf{Z}_{i,j},X_{i,j},\Delta_{i,j}^*,Y_{i,j})$: (1) the age of individual $i$ was incremented to compute the (partially) time-varying bivariate covariate vector $\textbf{Z}_{i,j}$; (2) serostatus was sampled according to \eqref{eq_p_seroconv} with $\{\alpha_s,\gamma_s,\tau_s\}=\{-1.6,(0.5,0.3),0.1\}$, $\{\alpha_x,\gamma_x,\tau_x\}=\{-3,(0.3,0.2),-0.1\}$, a complementary log-log link function, and linear calendar time trends $\textbf{b}_s(j)=\textbf{b}_x(j)=j$; (3) the variable $\Delta_{i,j}^*$ (the probability of receiving an HIV test) was sampled according to \eqref{eq_f_deltastar} with $\{\alpha_\delta,\gamma_\delta,\tau_\delta\}=\{-0.8,(0.2,0.1),0\}$ (with $\tau_\delta=0$ indicating that no calendar time trend was used) and a complementary log-log link function; (4) the outcome $Y_{i,j}$ was sampled according to \eqref{eq_p_outcome}, with $\{\alpha_y,\gamma_y,\tau_y,\beta\}=\{-3.5,(0.2,0.1),-0.1,0.4\}$ and $\textbf{b}_y(j)=j$. This process was terminated if an event occurred or if the end of the observation window ($j=20$) was reached. At this point, the vector $\Delta_i$ was calculated according to the function $g_\delta$ given in \eqref{eq_g_delta} and the vector $\textbf{U}_i$ was set to $\textbf{X}_i\Delta_i$. The vector $\textbf{X}_i$ was then removed from the dataset.

The likelihood function given in \eqref{eq_likelihood_miss} was programmed in R version 4.3.2. For each simulated dataset, the \texttt{optim()} function was used to numerically maximize and differentiate the likelihood to estimate the parameter vector, and the \texttt{hessian()} function of the \texttt{numDeriv} package was used to estimate the Hessian matrix. Performance was evaluated by estimating bias, standard errors, and 95\% confidence interval coverage. Simulations were conducted in R 4.3.2 and structured using the \texttt{SimEngine} simulation framework \citep{kenny2024simengine}; simulation code is available at \if1\anon{\url{https://github.com/Avi-Kenny/Discrete-time-survival-interval-censoring}}\else{[Github URL redacted]}\fi. Results based on $1,000$ simulation replicates are shown in Table \ref{table_sim_results} for a selection of model parameters; results were similar for other parameters and are suppressed for brevity. As expected with a correctly-specified parametric model, estimates are accurate overall, and minor deviations from expected operating characteristic values are likely due to a combination of finite sample bias and Monte Carlo error.

\begin{table}[ht!]
\begin{tabular}{l|l|l|l|l|l|l|l|}
\cline{2-8}
& $\beta$ & $\alpha_y$ & $\gamma_y[1]$ & $\tau_y$ & $\alpha_x$ & $\gamma_x[1]$ & $\tau_x$ \\ \hline
\multicolumn{1}{|l|}{True parameter value} & 0.4 & -3.5 & 0.2 & -0.1 & -3.0 & 0.3 & -0.1 \\ \hline
\multicolumn{1}{|l|}{Average estimate} & 0.409 & -3.542 & 0.214 & -0.086 & -2.974 & 0.294 & -0.109 \\ \hline
\multicolumn{1}{|l|}{Bias (absolute)} & 0.009 & -0.042 & 0.014 & 0.014 & 0.026 & -0.006 & -0.009 \\ \hline
\multicolumn{1}{|l|}{Average estimated standard error} & 0.149 & 0.169 & 0.091 & 0.058 & 0.36 & 0.176 & 0.129 \\ \hline
\multicolumn{1}{|l|}{Empirical standard error} & 0.143 & 0.157 & 0.085 & 0.058 & 0.305 & 0.163 & 0.117 \\ \hline
\multicolumn{1}{|l|}{95\% CI coverage} & 96.6\% & 96.1\% & 96.0\% & 93.9\% & 96.6\% & 95.5\% & 95.6\% \\ \hline
\end{tabular}
\captionsetup{width=0.9\linewidth}
\caption{Simulation results for a selection of model parameters across 1,000 simulation replicates, including bias, standard error estimation, and confidence interval coverage.}
\label{table_sim_results}
\end{table}

\section{Data Analysis}\label{sec_analysis}

\subsection{Dataset description and model specification}

We applied the methods developed in this paper to data from the Population Intervention Programme (formerly called the Africa Centre Demographic Information System), a large population-based open cohort in South Africa that has been followed since 2000 by the Africa Health Research Institute (AHRI). The purpose of establishing the cohort was ``to describe the demographic, social and health impacts of a rapidly progressing HIV epidemic in rural South Africa, and to monitor the impact of intervention strategies'' \citep{gareta2021cohort}. The cohort involves multiple data sources, including regular household surveys every four to six months to collect information on demographics and health outcomes, HIV serological testing as part of annual individual health surveys, and linkage to individual clinical records at local health facilities. For a thorough description of this cohort, see \cite{gareta2021cohort}. Our secondary analysis was conducted in concordance with the terms of the AHRI data use agreement, ensuring ethical use, confidentiality, and compliance with original institutional review board (IRB) guidelines.

The original dataset contained 2,799,721 person-years of data from 253,960 individuals between 2000 and 2022. We created an analysis subcohort by filtering this data to include only person-time units between years 2010 and 2022, between ages 13 and 60, and those residing within the original cohort study area (excluding those enrolled in a 2016 expansion of the study area) to ensure a consistent population. A small number ($<1\%$) of person-time units were dropped due to miscellaneous data quality issues. The analysis subcohort contained 879,098 person-years of data from 102,784 individuals (an average of 8.6 observation years per person), 106,695 HIV tests (an average of 1.0 test per person), and 5,531 deaths. 58\% individuals in the dataset never received an HIV test, 28\% had only negative tests, 13\% had only positive tests, and 1\% had a negative test followed by a positive test; Of those who received at least one test, the average number of tests was 2.5.

\begin{figure}[ht!]
    \centering
    \includegraphics[width=0.9\linewidth]{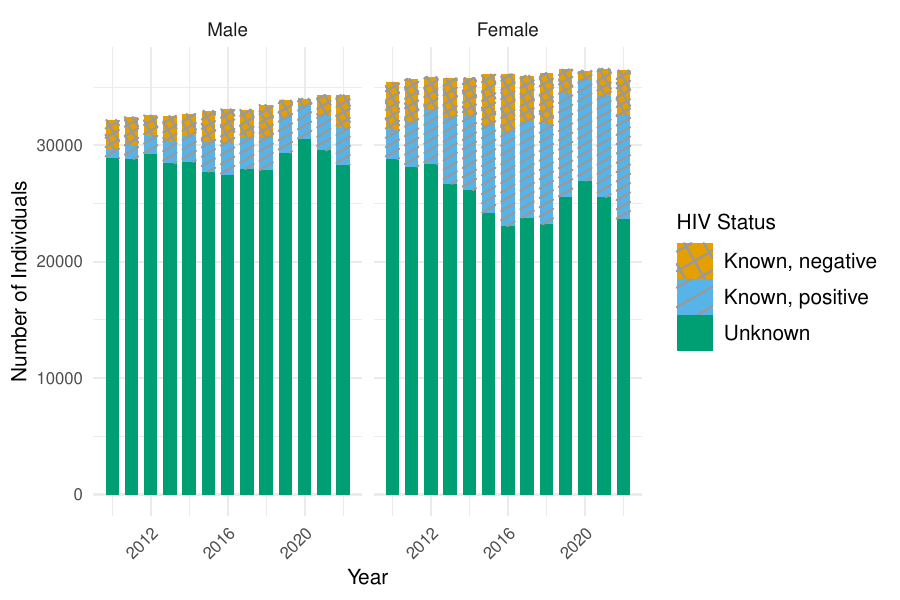}
    \captionsetup{width=0.9\linewidth}
    \caption{Number of individuals in the dataset by year, stratified by sex and HIV status (known positive, known negative, or unknown)}
    \label{fig_eda_1}
\end{figure}

Figure \ref{fig_eda_1} displays the total number of individuals in the dataset each year, stratified by sex and whether their HIV status is known positive, known negative, or unknown. Note that whether an individual's status is considered ``known'' at a given point in time is from the point of view of a researcher with access to that individual's full testing history. This figure highlights the need for an approach to modeling HIV status, since use of one of the ad-hoc methods described in the section \ref{sec-intro} would lead to a large proportion of available person-time being discarded.

As described in section \ref{ssec_model_spec}, a model specification involves selecting parametric forms for the model components. We fit the model separately in males and in females. For the discrete hazard of seroconversion, we included a linear term for calendar time and a spline term for age (this spline term and all others used in this model were natural cubic splines with four degrees of freedom, with one exception; see subsection \ref{ssec_data_notes} below). For the initial HIV status model component, we similarly included a linear term for calendar time, a spline term for age. For the discrete hazard of death, we included a spline term for age, a spline term for calendar time, and a term representing the hazard ratio of HIV-positive status relative to HIV-negative status as a function of age and calendar time (via a spline for age, a linear term for calendar time, and an interaction between the age and calendar time terms; this allowed for age-specific linear changes in this hazard ratio over calendar time).

\subsection{Data analysis results}

\begin{figure}[ht!]
    \centering
    \includegraphics[width=0.9\linewidth]{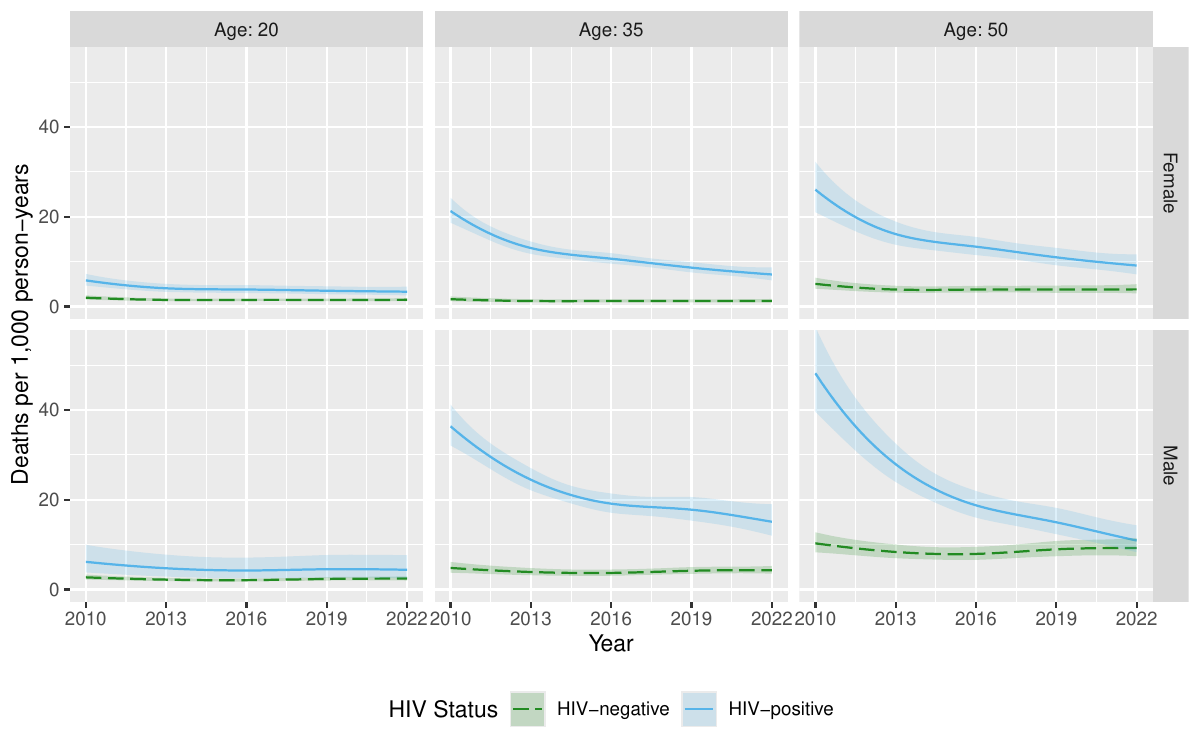}
    \captionsetup{width=0.9\linewidth}
    \caption{Conditional mortality rates (number of deaths per 1,000 person-years) by calendar time and HIV status, shown for several combinations of age and sex.}
    \label{fig_data_mortality_cal}
\end{figure}

Figure \ref{fig_data_mortality_cal} shows modeled mortality rates as a function of calendar time for several combinations of age and sex, stratified by HIV status, along with pointwise confidence intervals. As described above, the model allows for the hazard ratio to vary flexibly as a function of both age and calendar time, and we see this reflected in this plot; the ratio of the death rate in HIV-positive individuals relative to HIV-negative individuals is much higher in the age 35 group than in the age 20 or age 50 group and decreases over time in all age groups (for both males and females). Similar patterns are observed in males and females, with death rates remaining relatively stable over time in the HIV-negative subpopulation but decreasing substantially over time in the HIV-positive population.

\begin{figure}[ht!]
    \centering
    \includegraphics[width=0.9\linewidth]{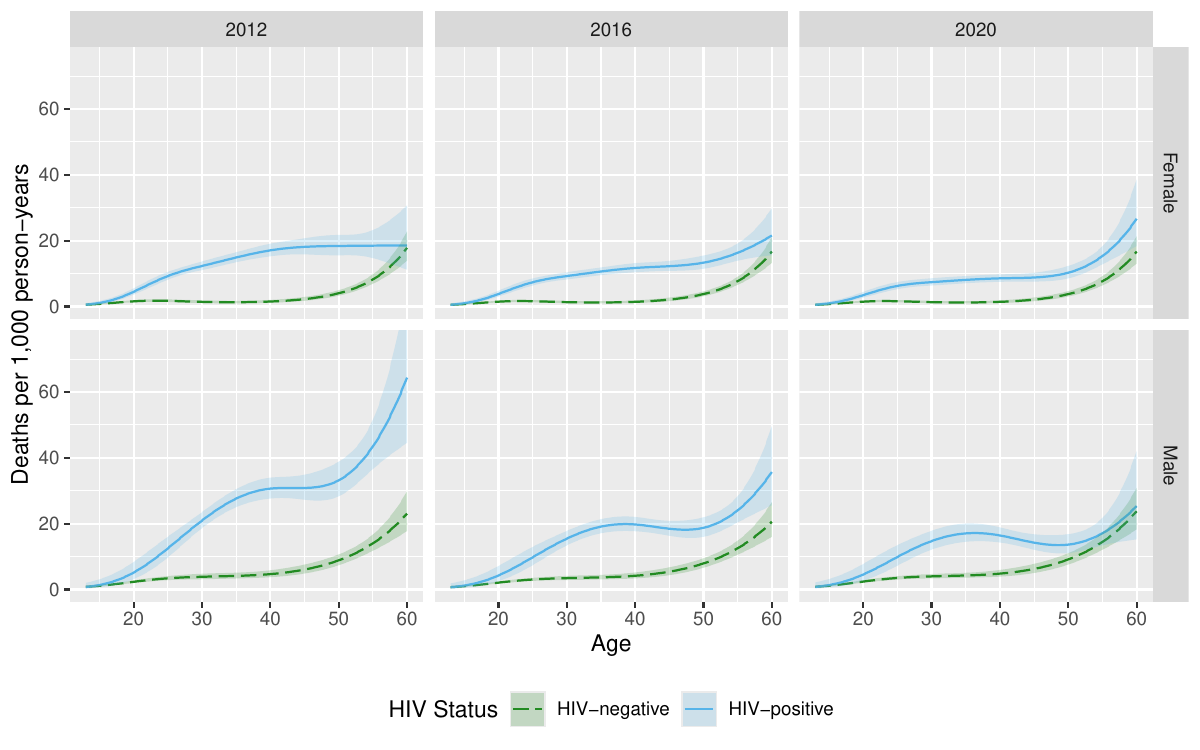}
    \captionsetup{width=0.9\linewidth}
    \caption{Conditional mortality rates (number of deaths per 1,000 person-years) by age and HIV status, shown for several combinations of calendar time and sex.}
    \label{fig_data_mortality_age}
\end{figure}

Figure \ref{fig_data_mortality_age} shows modeled mortality rates as a function of age for several combinations of calendar year and sex, stratified by HIV status, along with pointwise confidence intervals. This is essentially a different view of the same estimates from Figure \ref{fig_data_mortality_cal}, but highlights how the probability of mortality changes as a function of age, as well as how this function shifts over time. Interestingly, we see that the modeled death rate in 2012 among HIV-positive individuals over age 50 is roughly three times as high in males relative to females, but the magnitude of this difference fades over calendar time.

\begin{figure}[ht!]
    \centering
    \includegraphics[width=0.9\linewidth]{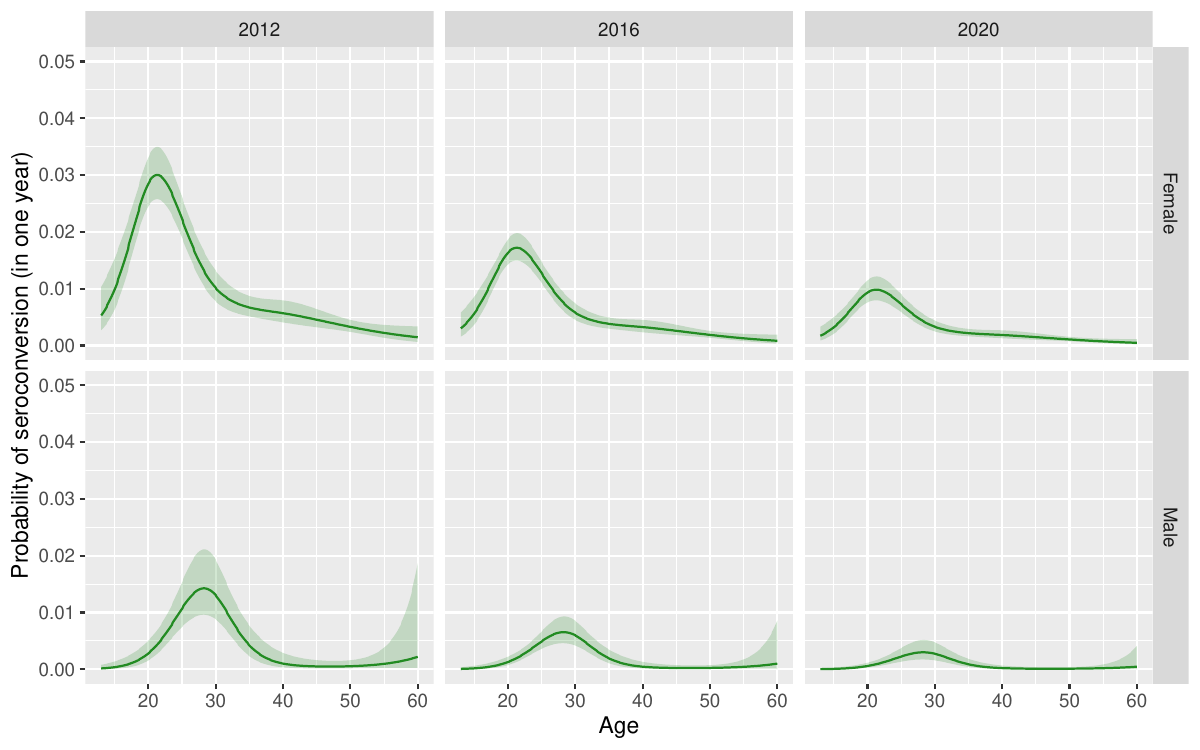}
    \captionsetup{width=0.9\linewidth}
    \caption{Conditional seroconversion discrete hazards (probability of seroconversion in one year) by age, shown for several combinations of calendar time and sex.}
    \label{fig_data_sero_age}
\end{figure}

Figure \ref{fig_data_sero_age} plots the discrete hazard of seroconversion (i.e., the probability that an individual will seroconvert in a single year) as a function of age for several combinations of calendar year and sex, along with pointwise confidence intervals. We observe that seroconversion probability peaks between the ages of 20 and 25 among females and peaks between the ages of 25 and 30 among males, and that these probabilities decrease substantially over time. These trends are consistent with previously-reported results among similar populations in South Africa \citep{vandormael2019declines,johnson2022effect}.

\begin{figure}[ht!]
    \centering
    \includegraphics[width=0.9\linewidth]{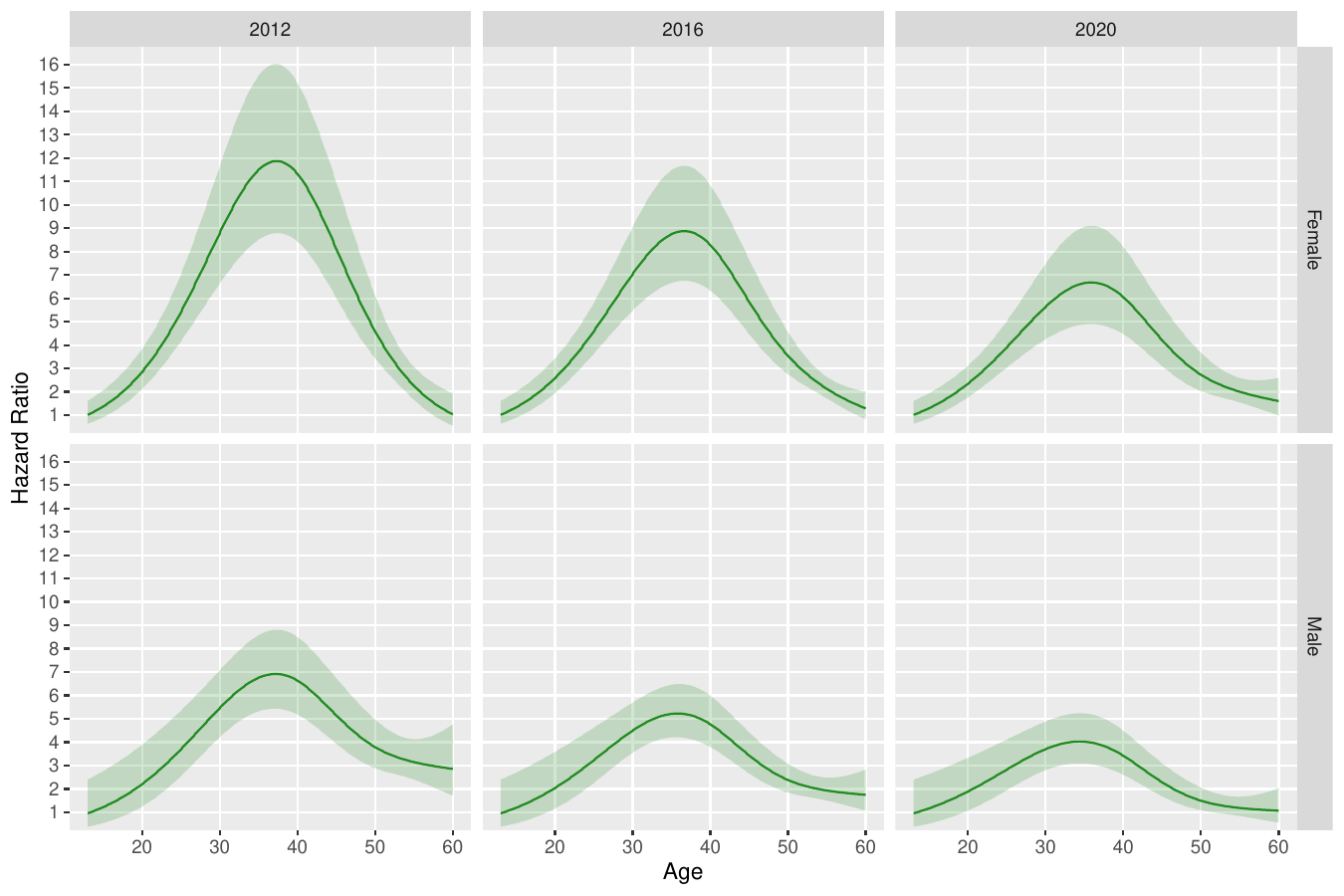}
    \captionsetup{width=0.9\linewidth}
    \caption{Hazard ratio for mortality, for HIV-positive status relative to HIV-negative status over calendar time.}
    \label{fig_HR_mort_hiv_age}
\end{figure}

Figure \ref{fig_HR_mort_hiv_age} shows the hazard ratio of HIV-positive individuals (relative to HIV-negative individuals) with respect to mortality risk as a function of age, along with pointwise confidence intervals, stratified by sex and displayed for 2012, 2016, and 2020. We observe that modeled hazard ratios are far lower for the youngest individuals, which is likely explained by the fact that HIV-related mortality often occurs many years after the transmission event. The hazard ratio peaks for both sexes between age 30 and 40, and then decreases substantially between ages 40 and 60, which is expected since there are many other competing causes of mortality for older individuals.

\begin{figure}[ht!]
    \centering
    \includegraphics[width=0.9\linewidth]{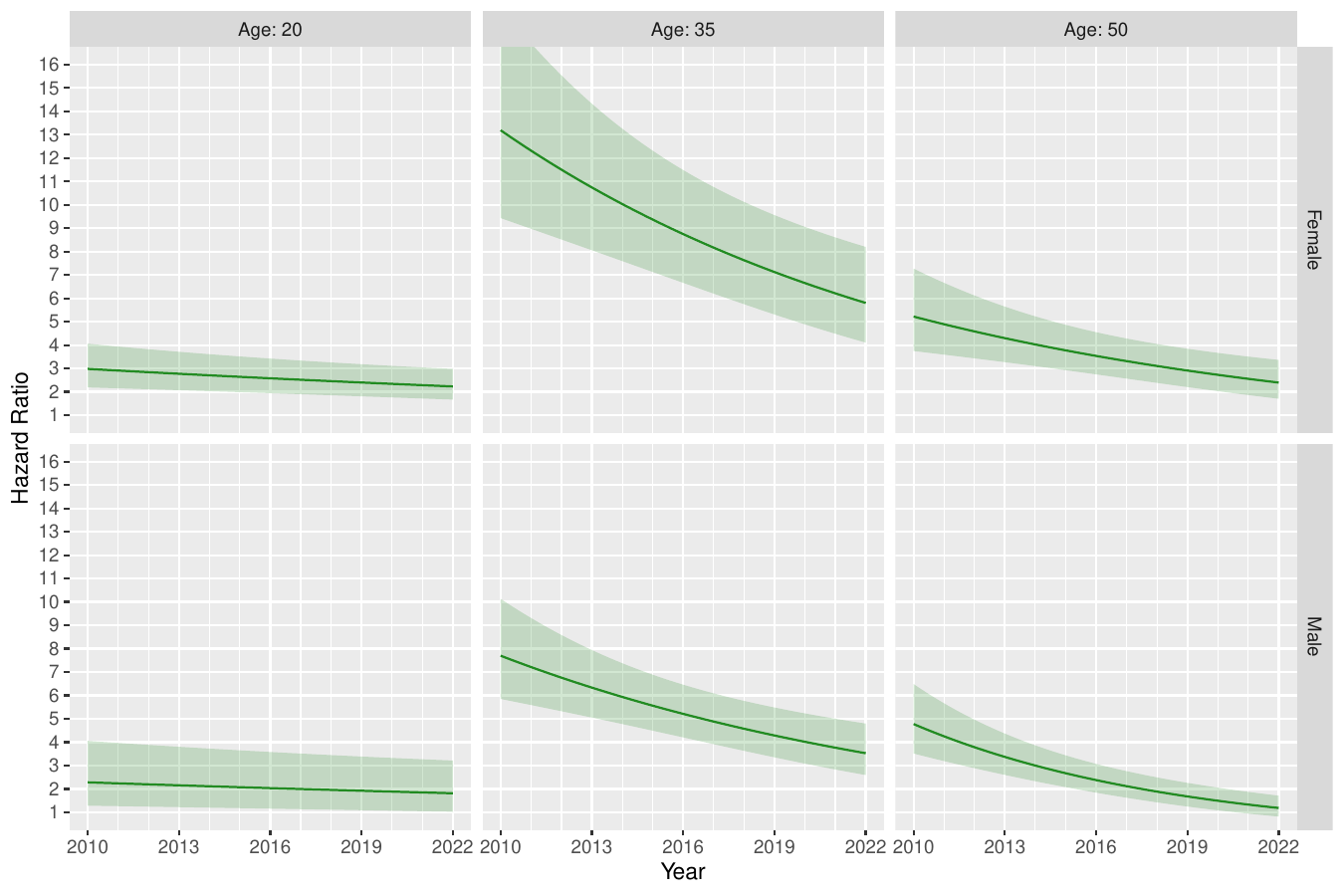}
    \captionsetup{width=0.9\linewidth}
    \caption{Hazard ratio for mortality, for HIV-positive status relative to HIV-negative status over calendar time.}
    \label{fig_HR_mort_hiv_year}
\end{figure}

Figure \ref{fig_HR_mort_hiv_year} is analogous to Figure \ref{fig_HR_mort_hiv_age}, but instead displays hazard ratio as a function of calendar time. This plot highlights the substantial decrease in the hazard ratio over calendar time, particularly among individuals in the middle of the age range. Similar patterns are observed in males and females.

\subsection{Technical notes}\label{ssec_data_notes}

One complication with dataset processing was how to integrate HIV testing data from prior to 2010 (the start of the analysis subcohort time window) if available; we chose to treat a positive HIV test from prior to 2010 as a positive HIV test in 2010, but to not ``carry forward'' information about negative tests; although perhaps not the most efficient approach, this choice should not lead to bias if the parametric form of the analysis model is correctly specified given the Markov assumptions.

The modeling results presented in this section were the result of several iterations of model development. The original model had numerical issues related to optimizer convergence and unstable Hessian estimates, and to tackle these issues, several simplifications were made, including conversion of several spline-based calendar time terms to linear terms. We chose to change the calendar time terms rather than the age terms since domain knowledge suggested that while (roughly) linear calendar time trends were plausible, nonlinear age-based relationships would be expected to hold (e.g., for the discrete hazard of seroconversion). Additionally, we had to remove one knot from the natural cubic spline for seroconversion (in the model for males only) due to convergence issues, which was done after diagnostic plots suggested that there were not enough seroconversion events among males under 20 to support the originally specified degree of flexibility. We also made several modifications to the model based on co-author discussion, most notably allowing the hazard ratio model (which was initially only allowed to vary with calendar time) to vary flexibly as a function of both calendar time and age. Inferential results should be interpreted accordingly.

\section{Discussion}\label{sec_discussion}

In this paper, we describe a discrete-time survival model that can be fit via maximum likelihood and used in applications involving an interval-censored covariate representing the occurrence of a secondary event that influences the conditional hazard of the primary event. In the context of the HIV serostatus application, the method proposed in this paper provides a way to fit a survival model using all available data on testing and outcomes. Historically, researchers have removed all data prior to the first HIV test (such that patients enter the risk set at the time of the first test) and/or all data at some point (e.g., two years) after the last HIV negative test. Both methods are ad-hoc and lead to a large proportion of observation time being discarded. The latter is particularly problematic for studies of long-term health outcomes (e.g., chronic disease incidence and mortality), because it leads to the majority of outcome data being discarded for individuals for whom the most recent HIV test is negative.

We consider the case in which there was a single secondary event. This implies that the conditional PMF given in \eqref{eq_f_uyd} could be computed (for a given individual) as a sum over $k+1$ terms, where $k$ is the number of time points measured for that individual. If there are $m$ secondary events, then the set $\mathcal{X}$ in \eqref{eq_f_uyd} over which the marginalization is done will be of size $(k+1)^m$, substantially affecting computation time, unless there is some sort of dependence structure between the events (e.g., one event cannot occur before the other). Exploration of the case involving multiple secondary events represents a worthwhile direction for future research, and will have immediate applications in similar HIV cohort studies for modeling the start of antiretroviral therapy.

Our approach relies on a proportional hazards structure in the specification of conditional discrete hazard functions for both the primary event and the secondary event, but allows for hazard ratios to vary arbitrarily as functions of any available baseline or time-varying covariates (as recommended in \citealp{hernan2010hazards}). Thus, many of the limitations and considerations that apply to proportional hazards models in general will apply here \citep{uno2014moving,royston2013restricted}. In general, one clear set of use cases for this model is settings for which a researcher has concluded based on scientific and statistical considerations that a Cox proportional hazards model is appropriate, but that involves the extra complication of an interval-censored covariate; however, a drawback of our model (unlike the Cox model) is that the baseline hazard of the outcome has to be specified parametrically.

It is possible to approach this problem by constructing an EM algorithm, as was done in \cite{goggins1999applying} and \cite{ahn2018cox}. One disadvantage of this approach is that it requires the user to integrate the full likelihood with respect to the density of the secondary event conditional on the primary event (and other covariates); while this is technically possible, it is difficult to specify this density in a principled manner that reflects the underlying causal mechanisms, since the occurrence of the secondary event is assumed to influence the hazard of the primary event, and not vice versa. A second disadvantage is that EM algorithms are known to often be computationally intense and require many iterations to achieve convergence \citep{ng2012algorithm}.

The findings from the data analysis in section \ref{sec_analysis} build on evidence from other studies of HIV incidence and risk of mortality among groups defined by HIV serostatus. In a study of the same population-based cohort, \cite{vandormael2019declines} estimated incidence using a series of surveys involving HIV testing and observed an overall decrease of 43\% between 2012 (0.040 seroconversion events per person-year) and 2017 (0.023 seroconversion events per person-year). Although incidence estimation is not the main goal of the model described in this paper, it is reassuring to see similar trends (see Figure \ref{fig_data_sero_age}), and an analysis allowing for more direct comparisons is a worthwhile future research direction. \cite{reniers2014mortality} also examined mortality rates by HIV status in a multi-country analysis, dealing with interval-censored serostatus by (1) assuming individuals who had a negative test followed by a positive test seroconvert at the midpoint between the two tests, and (2) censoring HIV-negative individuals after a certain (age and sex specific) period of time following their last negative test, and (3) censoring all individuals prior to their first HIV test. In their South Africa site in 2010, they estimate a mortality rate (per 1,000 person-years) of roughly 70 among HIV-positive males, 25 among HIV-positive females, 10 among HIV-negative males, and 5 among HIV-negative females, yielding sex-specific hazard ratios of roughly 7 and 5 among males and females, respectively. Our results are not directly comparable, as we allowed the hazard ratio to vary with age in addition to calendar time and person-time inclusion was different, but we see similar mortality rate patterns and very high hazard ratios, particularly among individuals aged 30 and 40. It is interesting to see that the hazard ratios peak in the same age range for both males and females; this trend might be largely driven by the fact that there are fewer competing causes of mortality for individuals in this age range in both sexes. Rough alignment of both the seroconversion model and the mortality model with existing estimates is encouraging.

One limitation of the approach taken here is that, in some applications, it will require researchers to ``artificially'' discretize the data, which involves both a coarsening of the data and the adding of many additional rows to the dataset. As the discretization grid becomes finer (i.e., as the time intervals shrink in length), the loss of information due to coarsening should decrease, eventually to the point of negligibility, but computation time will increase. Thus, there is a trade-off between computation time and precision that will need to be assessed in each application individually; this is a feature common to all discrete survival models. Additionally, putting datasets into a longer format in which individuals contribute multiple rows of person-time often has to be done anyways, as one would do when fitting a Cox model with time-varying covariates. A full discussion of the relative merits of continuous-time versus discrete-time survival models is outside the scope of this work; see, for example, \cite{suresh2022survival}.

A second limitation of this approach is that it assumes the conditional hazard of the primary event of interest depends only on whether or not the secondary event occurred, rather than the time since its occurrence. In cases in which no individuals have experienced the secondary event at the start of the study, this represents a straightforward extension to the current work, since the ``time since secondary event'' variable can be computed deterministically for each element of the set $\mathcal{X}$ in \eqref{eq_f_uyd}. However, if for some individuals, the event can occur before the start of the observation window (as is the case in the HIV motivating dataset), this extension becomes more difficult, as one must posit a model for the conditional distribution of the ``time since secondary event'' variable at the start of the observation window, the parameters of which may be difficult or impossible to fit for a given dataset without making strong assumptions. If this assumption is violated, the hazard ratio for the secondary event will represent a weighted average of the time-specific (time since secondary event) hazard ratios. Note that this issue is discussed in \cite{risher2021age} and a potential solution is proposed.

A third limitation is that this model assumes a full parametric likelihood for the observed data and the missingness mechanism; if any components of the model are misspecified, estimators may be biased and inference will be invalid; this is a limitation of parametric models in general. In particular, we are assuming that, conditional on covariates, individuals for whom we have no testing data have the same hazard of seroconversion as individuals for whom we do have testing data. This assumption may not be true in practice, and one may wish to run a model that excludes those who have never received an HIV test, which can be viewed either as a sensitivity analysis or as an analysis making inference about a restricted target population. In general, the dynamics of HIV epidemiology are complex and change over time, and any model will be a simplification of the true data-generating mechanism. Future work can also explore more complex parametric specifications (such as including age-by-time interactions in the seroconversion and mortality hazard models), the use of different missingness mechanisms, and so on.

\section{Disclosure statement}\label{disclosure-statement}

The authors do not have any conflicts of interest to declare.

\section{Data Availability Statement}\label{data-availability-statement}

Data are available upon reasonable request by accessing https://data.ahri.org/index.php/catalog/1183 via the Africa Health Research Institute Data Repository. Further information about reproducing results is given in the Author Contributions Checklist (ACC) form.

\section*{Acknowledgments}

Research reported in this publication was supported by the National Institute Of Allergy And Infectious Diseases of the National Institutes of Health [Award Number \if1\anon{R37-AI029168}\else{redacted}\fi], the National Heart, Lung, and Blood Institute of the National Institutes of Health [Award number \if1\anon{K24-HL166024}\else{redacted}\fi], and the Wellcome Trust [Award number Wellcome Strategic Core award: \if1\anon{227167/A/23/Z}\else{redacted}\fi]. The funders had no role in study design, analysis, and interpretation of data; in the writing of the manuscript; or in the decision to submit the manuscript for publication. The content is solely the responsibility of the authors and does not necessarily represent the official views of the National Institutes of Health.

\bibliography{bibliography.bib}

\end{document}